\documentclass{JAC2003}
\usepackage{graphicx}
\usepackage{booktabs}
\newtheorem{proposition}{Proposition}
\usepackage{epsfig}
\usepackage{color}

\begin{document}

\title{RELATIONS BETWEEN PROJECTED EMITTANCES AND EIGENEMITTANCES
\vspace{-0.5cm}}

\author{V.Balandin\thanks{vladimir.balandin@desy.de}, 
W.Decking, N.Golubeva \\
DESY, Hamburg, Germany}

\maketitle

\begin{abstract}
We give necessary and sufficient conditions
that two sets of positive real numbers must satisfy
in order to be realizable as eigenemittances and projected
emittances of a beam matrix.
The information provided by these conditions 
sets limits on what one can to achieve
when designing a beam line to perform advanced emittance manipulations.
\end{abstract}

\vspace{-0.1cm}
\section{Introduction}

\vspace{-0.1cm}
Projected emittances are quantities which are used to characterize
transverse and longitudinal beam dimensions in the laboratory 
coordinate system and are invariants under linear uncoupled (with respect 
to the laboratory coordinate system) symplectic transport.
Eigenemittances are quantities which give beam dimensions in the
coordinate frame in which the beam matrix is uncoupled between degrees of
freedom and are invariants under arbitrary (possibly coupled)
linear symplectic transformations. If the beam matrix is uncoupled
already in the laboratory frame,
then the set of projected emittances coincides with the set of 
eigenemittances, and if the beam matrix has correlations between
different degrees of freedom, then these two sets are different.
This fact, though looking simple, has interesting applications in accelerator
physics and gives the theoretical basis for the round-to-flat 
transformation of angular momentum dominated beams invented 
by Derbenev~\cite{Derbenev01}.
In his scheme the beam with equal transverse projected
emittances (round beam) but with nonequal eigenemittances 
is first produced in an axial magnetic field. Then 
the correlations in the beam matrix are removed by a 
downstream set of skew quadrupoles and projected emittances
become equal to the eigenemittances, which means that the beam 
transverse dimensions become different from each other.

This work and further development of the advanced emittance manipulation
techniques (see, for example \cite{Derbenev02,Kim,Carl} 
and references therein) naturally raise the following
question: what are the relations between projected emittances
and eigenemittances?
As concerning already known results, in general situation they
are limited to the so-called classical uncertainty principle, which
states that none of projected emittances can be smaller than the
minimal eigenemittance (see, for example, ~\cite{Dragt}).
Besides that, in the specific two degrees of freedom case, a number
of useful results can be found in ~\cite{Brown}.

The purpose of this article is to give
the necessary and sufficient conditions
which two sets of positive real numbers must satisfy
in two and three degrees of freedom cases
in order to be realizable as eigenemittances and projected
emittances of a beam matrix.

\vspace{-0.1cm}
\section{Beam matrix and emittances}

\vspace{-0.1cm}
Let us consider a collection of points in $2n$-dimensional 
phase space (a particle beam) and let, for each particle, 
 
\vspace{-0.2cm}
\noindent
\begin{eqnarray}
z \,=\,(q_1, \, p_1, \, \ldots, \, q_n, \, p_n)^{\top}
\label{intr_1}
\end{eqnarray}

\vspace{-0.2cm}
\noindent
be a vector of canonical coordinates $q_m$ and momenta $p_m$.
Then, as usual, the beam (covariance) matrix is
defined as

\vspace{-0.2cm}
\noindent
\begin{eqnarray}
\Sigma =
\left\langle
\left(z - \langle z \rangle \right) 
\cdot 
\left(z - \langle z \rangle \right)^{\top}
\right\rangle,
\label{intr_2}
\end{eqnarray}

\vspace{-0.2cm}
\noindent
where the brackets $\langle \, \cdot \, \rangle$
denote an average over a distribution of the particles in the beam.
By definition, the beam matrix $\Sigma$ is symmetric positive semidefinite and 
in the following we will restrict our considerations to the situation
when this matrix is nondegenerated and therefore positive definite. 
For simplification of notations and without loss of generality,
we will also assume that the beam has vanishing
first-order moments, i.e. $\big< z \big> \,=\, 0$.

Let $s$ be the independent variable and let $T = T(\tau)$ 
be the nondegenerated matrix which
propagates particle coordinates from the state $s = 0$ to the
state $s = \tau$, i.e let

\vspace{-0.2cm}
\noindent
\begin{eqnarray}
z(\tau) \,=\, T\, z(0).
\label{intr_4}
\end{eqnarray}

\vspace{-0.2cm}
\noindent
Then from (\ref{intr_2}) and (\ref{intr_4}) it follows
that the matrix $\Sigma$ evolves between these two
states according to the congruence

\vspace{-0.2cm}
\noindent
\begin{eqnarray}
\Sigma(\tau) \,=\, T\, \Sigma(0)\, T^{\top}.
\label{intr_5}
\end{eqnarray}

\vspace{-0.2cm}
Let us write the $2n \times 2n$ matrix $\Sigma$ in block-matrix form

\vspace{-0.2cm}
\noindent
\begin{eqnarray}
\Sigma \;=\;
\left(
\begin{array}{cccc}
\Sigma_{11} & \Sigma_{12} & \cdots & \Sigma_{1n} \\
\vdots      & \vdots      & \ddots & \vdots      \\
\Sigma_{n1} & \Sigma_{n2} & \cdots & \Sigma_{nn}
\end{array}
\right),
\label{intr_3}
\end{eqnarray}

\vspace{-0.2cm}
\noindent
where the entries $\Sigma_{mk}$ are $2 \times 2$ matrices.
Because $\Sigma$ is symmetric, the blocks satisfy the relations
$\Sigma_{mk} = \Sigma_{km}^{\top}$ for all $m, k = 1, \ldots, n$.
One says that the beam matrix $\Sigma$ is uncoupled if all its $2 \times 2$ 
blocks $\Sigma_{mk}$ with $m \neq k$ are equal to zero,
and one says that the $m$-th degree of freedom
in the beam matrix $\Sigma$ is decoupled from the others if
$\Sigma_{mk} = \Sigma_{km} = 0$ for all $k \neq m$. 

If, similar to the matrix $\Sigma$, we will partition the matrix $T$ 
into submatrices $T_{mk}$,
then one can rewrite the transport equation (\ref{intr_5})  
in the form of a system involving only $2 \times 2$ submatrices
of the matrices $\Sigma$ and $T$

\vspace{-0.2cm}
\noindent
\begin{eqnarray}
\Sigma_{mk}(\tau)\,=\,\sum\limits_{l, p = 1}^{n} T_{ml}\,
\Sigma_{lp}(0)\, T_{kp}^{\top},
\quad
m,k = 1,\ldots,n.
\label{intr_18}
\end{eqnarray}

\vspace{-0.2cm}
In analogy with the matrix $\Sigma$, one says that
the transport matrix $T$ is uncoupled if all its
blocks $T_{mk}$ with $m \neq k$ are equal to zero,
and one says that the $m$-th degree of freedom
in the transport matrix $T$ is decoupled from the others if
$T_{mk} = T_{km} = 0$ for all $k \neq m$. 

In the following we will assume that the beam transport matrix $T$ is 
symplectic, which is equivalent to say that it satisfies the relations

\vspace{-0.2cm}
\noindent
\begin{eqnarray}
T \, J_{2n} \,T^{\top} \,=\,
T^{\top} J_{2n} \,T \,=\, J_{2n}
\label{intr_8}
\end{eqnarray}

\vspace{-0.3cm}
\noindent
where

\vspace{-0.3cm}
\noindent
\begin{eqnarray}
J_{2n} \;=\;\mbox{diag}
\Bigg(
\underbrace{
\left(
\begin{array}{rr}
0 & 1\\
-1 & 0
\end{array}
\right),
\ldots,
\left(
\begin{array}{rr}
0 & 1\\
-1 & 0
\end{array}
\right)
}_{n}
\Bigg)
\label{intr_7}
\end{eqnarray}

\vspace{-0.2cm}
\noindent
is the $2n \times 2n$  symplectic unit matrix.

\begin{figure}[t]
    \centering
    \includegraphics*[width=75mm]{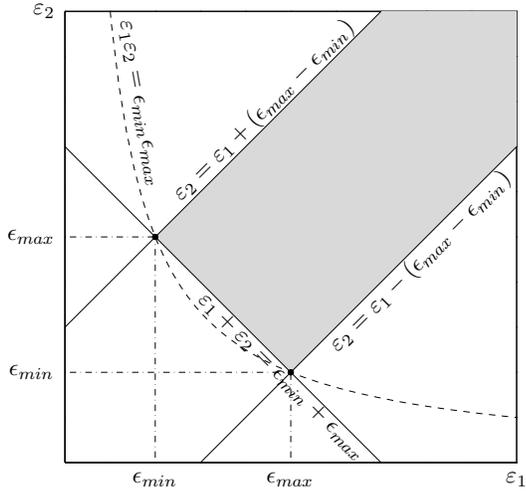}
    \vspace{-0.5cm}
    \caption{Shaded area shows all possible values
    of projected emittances $\varepsilon_1$ and $\varepsilon_2$
    of a $4 \times 4$ beam matrix $\Sigma$ with fixed 
    eigenemittances $\epsilon_{min}$ and $\epsilon_{max}$.
    If $\epsilon_{min} = \epsilon_{max}$, then the shaded half-strip
    turns into a ray (half-line).}
    \label{fig1}
\vspace{-0.3cm}
\end{figure}

Using partitioning into $2 \times 2$ submatrices
the two (equivalent) conditions for 
the matrix $T$ to be symplectic (\ref{intr_8}) can be rewritten in 
the form of the following set of equations:

\vspace{-0.2cm}
\noindent
\begin{eqnarray}
\sum\limits_{l = 1}^{n} \,T_{ml}\, J_2\, T_{kl}^{\top} 
\;=\;
\sum\limits_{l = 1}^{n} \,T_{lm}^{\top}\, J_2\, T_{lk} 
\;=\; \delta_{mk} \,J_2, 
\label{intr_9}
\end{eqnarray}

\vspace{-0.2cm}
\noindent
where $m, k = 1, \ldots, n$ and $\delta_{mk}$ is Kronecker's delta.

Because for an arbitrary $2 \times 2$ matrix $X$ 

\vspace{-0.2cm}
\noindent
\begin{eqnarray}
X\,J_2\,X^{\top} \,=\,
X^{\top} J_2\,X \,=\, \det(X) \cdot J_2,
\label{intr_10}
\end{eqnarray}

\vspace{-0.2cm}
\noindent
the equations (\ref{intr_9}) give us the following
important identities

\vspace{-0.2cm}
\noindent
\begin{eqnarray}
\sum\limits_{l = 1}^{n} \det(T_{ml}) \,=\,
\sum\limits_{l = 1}^{n} \det(T_{lm})
\,=\, 1,
\label{intr_11}
\end{eqnarray}

\vspace{-0.2cm}
\noindent
which are valid for all $m = 1, \ldots, n$.

Projected emittances $\varepsilon_m$
are the rms phase space areas covered by projections
of the particle beam onto each coordinate plane $(q_m, p_m)$

\vspace{-0.2cm}
\noindent
\begin{eqnarray}
\varepsilon_m 
= {\det}^{1/2}(\Sigma_{mm})
= \sqrt{\langle q_m^2 \rangle \langle p_m^2 \rangle -
\langle q_m p_m \rangle^2}.
\label{Sec2C02}
\end{eqnarray}

\vspace{-0.2cm}
Let us assume that in the matrix $T$
the $m$-th degree of freedom is decoupled from the others.
Then from equations (\ref{intr_18}) one obtains that

\vspace{-0.2cm}
\noindent
\begin{eqnarray}
\Sigma_{mm}(\tau)\,=\, T_{mm}\,
\Sigma_{mm}(0)\, T_{mm}^{\top},
\label{intr_19}
\end{eqnarray}

\vspace{-0.2cm}
\noindent
and because due to (\ref{intr_11}) the submatrix
$T_{mm}$ has unit determinant, we see that the
projected emittance $\varepsilon_m$ is conserved
during the beam transport independently if
the $m$-th degree of freedom in the matrix $\Sigma(0)$
is decoupled from the others or not.

\begin{figure}[t]
    \centering
    \includegraphics*[width=75mm]{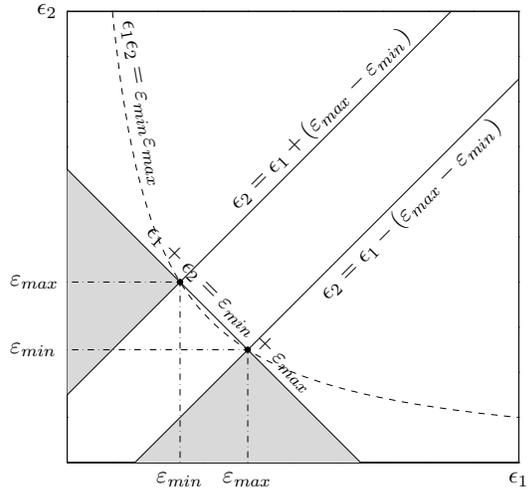}
    \vspace{-0.5cm}
    \caption{Shaded area (with exception of axes
    $\epsilon_1 = 0$ and $\epsilon_2 = 0$) shows all possible values
    of eigenemittances $\epsilon_1$ and $\epsilon_2$
    of a $4 \times 4$ beam matrix $\Sigma$ with fixed 
    projected emittances $\varepsilon_{min}$ and $\varepsilon_{max}$.
    If $\varepsilon_{min} = \varepsilon_{max}$, then two shaded
    triangles merge into one.}
    \label{fig2}
\vspace{-0.3cm}
\end{figure}

Using symplecticity of the transport matrix $T$
the congruence (\ref{intr_5}) can be transformed
into the following equivalent form

\vspace{-0.2cm}
\noindent
\begin{eqnarray}
(\Sigma J_{2n}) (\tau)\;=\; T \cdot (\Sigma J_{2n})(0)\cdot T^{-1}.
\label{vectZ_7}
\end{eqnarray}

\vspace{-0.2cm}
\noindent
From this form of the equation (\ref{intr_5}) we see that the eigenvalues 
of the matrix $\Sigma J_{2n}$ are invariants, because (\ref{vectZ_7}) is a similarity 
transformation. The matrix $\Sigma J_{2n}$ is nondegenerated and is similar to 
the skew symmetric matrix $\,\Sigma^{1/2} J_{2n} \,\Sigma^{1/2}$

\vspace{-0.2cm}
\noindent
\begin{eqnarray}
\Sigma J_{2n}\;=\; \Sigma^{1/2} \cdot 
(\Sigma^{1/2} J_{2n} \,\Sigma^{1/2}) \cdot \Sigma^{-1/2},
\label{vectZ_8}
\end{eqnarray}

\vspace{-0.2cm}
\noindent
which means that its spectrum is of the form

\vspace{-0.2cm}
\noindent
\begin{eqnarray}
\pm i \epsilon_1, \,\ldots, \,\pm i \epsilon_n,
\label{vectZ_9}
\end{eqnarray}

\vspace{-0.2cm}
\noindent
where all $\epsilon_m > 0$ and $i$ is the imaginary unit.
The quantities $\epsilon_m$ are called eigenemittances 
and generalize the property of the projected emittances
to be invariants of uncoupled beam transport
to the fully coupled case ~\cite{MomInv02}.

The other approach to the concept of eigenemittances is the way pointed out 
by Williamson's theorem (see, for example, references in ~\cite{MomInv02}).
This theorem tells us that one can diagonalize any positive definite 
symmetric matrix $\Sigma$ by congruence using a symplectic matrix $M$

\vspace{-0.2cm}
\noindent
\begin{eqnarray}
M \,\Sigma \, M^{\top} \;=\; D,
\label{A2_1}
\end{eqnarray}

\vspace{-0.2cm}
\noindent
and that the diagonal matrix $D$ has the very simple form

\vspace{-0.2cm}
\noindent
\begin{eqnarray}
D\,=\,\mbox{diag}(\Lambda, \Lambda),
\;\;
\Lambda \,=\, \mbox{diag} (\epsilon_1, \, \ldots, \, \epsilon_ n) > 0,
\label{A2_2}
\end{eqnarray}

\vspace{-0.2cm}
\noindent
where the diagonal elements $\epsilon_m$ are the moduli of the eigenvalues 
of the matrix $\Sigma J_{2n}$. The matrix $M$ in (\ref{A2_1}) is not unique, but 
the diagonal entries of the Williamson's normal form $D$ (eigenemittances) 
are unique up to a reordering.

\begin{figure*}[t]
\centerline{
\hbox{\psfig{figure=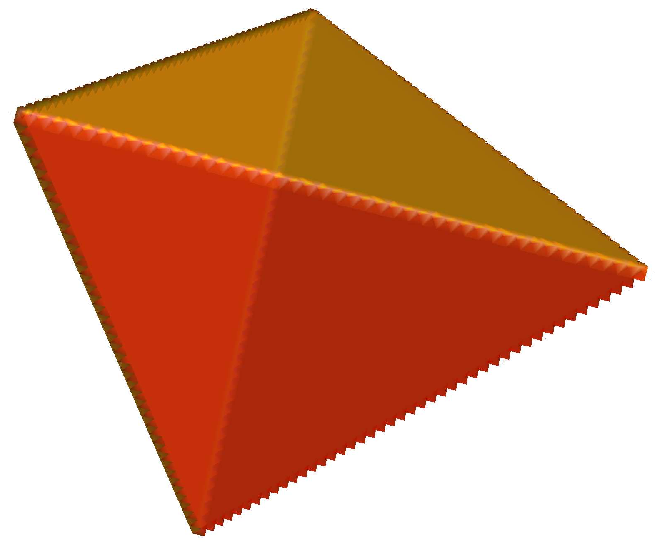,angle=0,width=4.0cm}}
\hbox{\psfig{figure=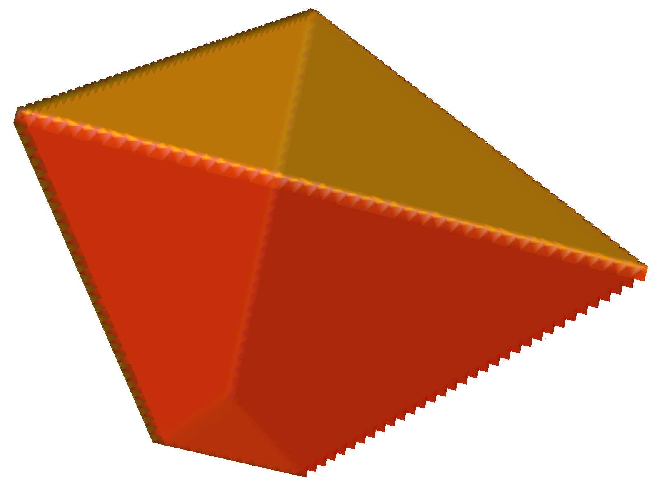,angle=0,width=4.0cm}}
\hbox{\psfig{figure=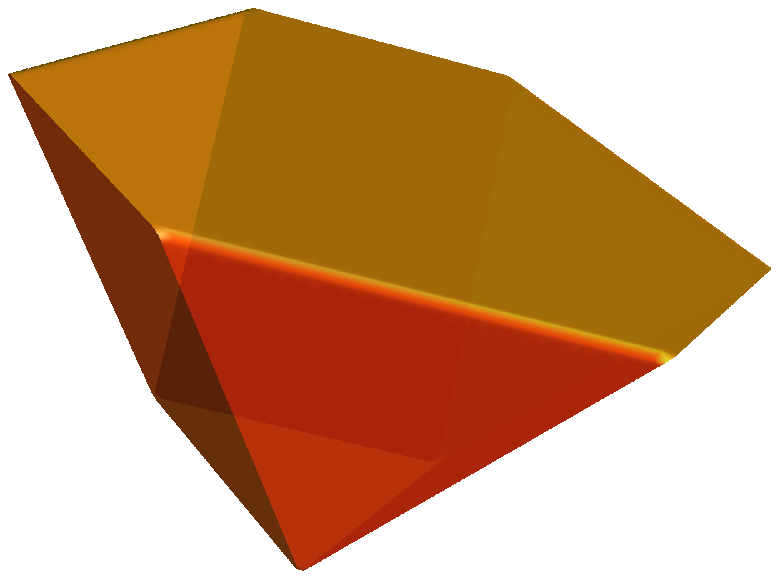,angle=0,width=4.0cm}}
\hbox{\psfig{figure=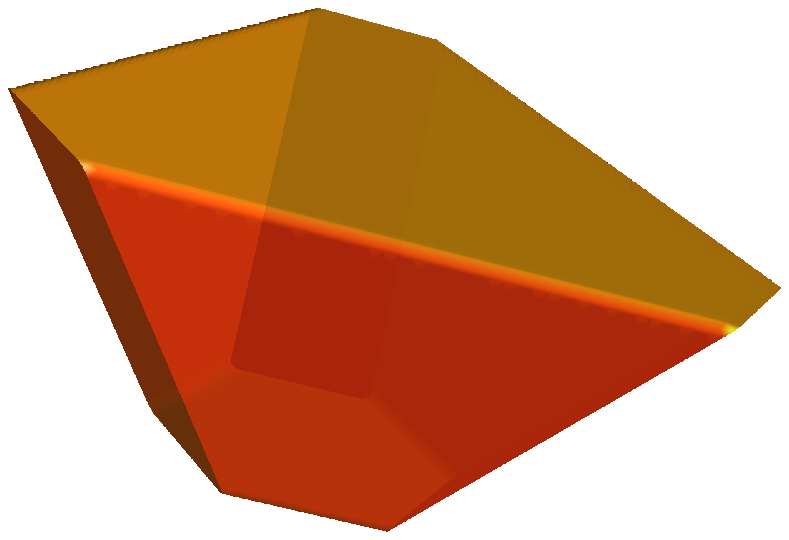,angle=0,width=4.0cm}}}
\vspace{-0.5cm}
\caption{Schematic drawing of area allowed for the projected emittances
of a $6 \times 6$ beam matrix $\Sigma$ with fixed eigenemittances. 
There are four geometrically distinguishable situations
which are shown from left to right and correspond to the relations
$\epsilon_{min}=\epsilon_{mid}=\epsilon_{max}$,
$\epsilon_{min}<\epsilon_{mid}=\epsilon_{max}$,
$\epsilon_{min}=\epsilon_{mid}<\epsilon_{max}$,
and
$\epsilon_{min}<\epsilon_{mid}<\epsilon_{max}$,
respectively.}
\vspace{-0.3cm}
\label{fig3}
\end{figure*}

It is clear that not only eigenemittances themselves, but also
an arbitrary function of them is an invariant. In particular, in the
following we will make use of invariants

\vspace{-0.1cm}
\noindent
\begin{eqnarray}
I_{2m} =
(-1)^m 
\mbox{tr}\left[(\Sigma J_{2n})^{2m}\right] / 2 
= \epsilon_1^{2m} + \ldots + \epsilon_n^{2m}.
\label{dd_2}
\end{eqnarray}

\section{Characterization of Uncoupled Beam Matrix
and Lower Bounds for Projected Emittances}

\vspace{-0.1cm}
In this section we summarize what it is possible to say about
beam matrix and its emittances in the arbitrary degrees of
freedom case. Note that not all presented relations are new.
For example, the inequality (\ref{prop2_1}) is the well known
classical uncertainty principle.
 
\vspace{-0.2cm}
\begin{proposition}

Let $\Sigma$ be a $2n \times 2n$ beam matrix and let
positive real numbers $\epsilon_1, \ldots, \epsilon_n$ and
$\varepsilon_1, \ldots, \varepsilon_n$ be its eigenemittances
and projected emittances, respectively. Then,
the following statements are equivalent:

\begin{itemize}

\vspace{-0.2cm}
\item[a)]
The beam matrix $\Sigma$ is uncoupled.

\vspace{-0.2cm}
\item[b)]
The set of projected emittances coincides with the set of eigenemittances.

\vspace{-0.2cm}
\item[c)]
The product of projected emittances is equal to the product of
eigenemittances

\vspace{-0.3cm}
\noindent
\begin{eqnarray}
\varepsilon_1 \, \varepsilon_2 \, \ldots \, \varepsilon_n
\, = \,
\epsilon_1    \, \epsilon_2    \, \ldots \, \epsilon_n.
\label{prop_1}
\end{eqnarray}

\end{itemize}

\end{proposition}

\vspace{-0.3cm}
\begin{proposition}

Let us denote by $\epsilon_{min}$ and $\varepsilon_{min}$ 
the minimums
of the quantities $\epsilon_1, \ldots , \epsilon_n$ and
$\varepsilon_1, \ldots , \varepsilon_n$ respectively.
Then, the following statements hold:

\vspace{-0.15cm}
\noindent
\begin{eqnarray}
\varepsilon_{min} \,\geq\,\epsilon_{min},
\label{prop2_1}
\end{eqnarray}

\vspace{-0.25cm}
\noindent
\begin{eqnarray}
\varepsilon_1 + \varepsilon_2 + \ldots + \varepsilon_n
\, \geq \,
\epsilon_1    + \epsilon_2    + \ldots + \epsilon_n,
\label{prop_2_2}
\end{eqnarray}

\vspace{-0.25cm}
\noindent
\begin{eqnarray}
\varepsilon_1 \, \varepsilon_2 \, \ldots \, \varepsilon_n
\, \geq \,
\epsilon_1    \, \epsilon_2    \, \ldots \, \epsilon_n.
\label{prop_2_3}
\end{eqnarray}

\end{proposition}

\vspace{-0.2cm}
\section{Two Degrees of Freedom}

In the two degrees of freedom case the eigenemittances
can be calculated according to the explicit formula

\vspace{-0.1cm}
\noindent
\begin{eqnarray}
\epsilon_{1,2} \,=\, 
\sqrt{\frac{I_2}{2} \,\pm\, 
\sqrt{\left(\frac{I_2}{2}\right)^2 \,-\, \det(\Sigma)}},
\label{vectZ_2_1}
\end{eqnarray}

\vspace{-0.1cm}
\noindent
and the exact relations between them and 
projected emittances are given by the following proposition:

\begin{proposition}

The positive real numbers $\epsilon_1$, $\epsilon_2$
and  $\varepsilon_1$, $\varepsilon_2$ can be realized as
eigenemittances and projected emittances of a $4 \times 4$
beam matrix $\Sigma$ if and only if 
the following two inequalities hold:

\noindent
\begin{eqnarray}
\left\{
\begin{array}{ccc}
\vspace{0.2cm}
\varepsilon_1 + \varepsilon_2   & \geq &  \epsilon_1 + \epsilon_2\\
|\,\varepsilon_1 - \varepsilon_2| & \leq & |\epsilon_1 - \epsilon_2|
\end{array}
\right.
\label{prop_3}
\end{eqnarray}

\end{proposition}

\newpage

The geometrical interpretation of the inequalities
(\ref{prop_3}) can be seen in Fig.1 and Fig.2.

\vspace{-0.2cm}
\section{Three Degrees of Freedom}

\vspace{-0.1cm}
In the three degrees of freedom case the eigenemittances
can be found as positive roots of the bicubic equation

\vspace{-0.2cm}
\noindent
\begin{eqnarray}
\epsilon^6 - I_2 \cdot \epsilon^4 +
(1/2) (I_2^2 - I_4) \cdot \epsilon^2 
- \det(\Sigma) = 0,
\label{ThreeDF_7}
\end{eqnarray}

\vspace{-0.2cm}
\noindent
and the exact relations between them and 
projected emittances are given by the following proposition:

\begin{proposition}

The positive real numbers  $\varepsilon_1$, $\varepsilon_2$, $\varepsilon_3$ and
$\epsilon_{min} \leq \epsilon_{mid} \leq \epsilon_{max}$ can be realized as
projected emittances and eigenemittances of a $6 \times 6$
beam matrix $\Sigma$ if and only if 
the following inequalities hold:

\noindent
\begin{eqnarray}
\left\{
\begin{array}{rcl}
\vspace{0.2cm}
\varepsilon_1 + \varepsilon_2 - \varepsilon_3  &\geq&
\epsilon_{min} + \epsilon_{mid} - \epsilon_{max}\\
\vspace{0.2cm}
\varepsilon_1 - \varepsilon_2 + \varepsilon_3  &\geq&
\epsilon_{min} + \epsilon_{mid} - \epsilon_{max}\\
\vspace{0.2cm}
-\varepsilon_1 + \varepsilon_2 + \varepsilon_3  &\geq&
\epsilon_{min} + \epsilon_{mid} - \epsilon_{max}\\
\vspace{0.2cm}
\varepsilon_1 + \varepsilon_2 + \varepsilon_3  &\geq&
\epsilon_{min} + \epsilon_{mid} + \epsilon_{max}\\
\vspace{0.2cm}
\varepsilon_{min} &\geq& \epsilon_{min}
\end{array}
\right.
\label{ThreeDF_1}
\end{eqnarray}

\end{proposition}

The geometrical interpretation of these inequalities
for the case when eigenemittances are fixed can be seen in Fig.3.

\end{document}